\documentclass[3p,twocolumn,sort&compress]{elsarticle}
\usepackage{amsmath, bm}
\usepackage{amssymb}
\usepackage{amsmath}
\usepackage{xcolor}
\usepackage{booktabs}
\usepackage{upquote}
\usepackage{listings}
\journal{Computer Physics Communications}

\usepackage{natbib}
\usepackage{hyperref}
\hypersetup{
        colorlinks=true,
}

\newcommand{\gv}[1]{\ensuremath{\mbox{\boldmath$ #1 $}}} 
\newcommand{\pd}[2]{\frac{\partial #1}{\partial #2}} 
\newcommand{\ket}[1]{\left| #1 \right \rangle} 
\newcommand{\braket}[2]{\left \langle #1 \vphantom{#2} \right| \left. #2 \vphantom{#1} \right \rangle} 
\newcommand{\matrixel}[3]{\left \langle #1 \vphantom{#2#3} \right| #2 \left| #3 \vphantom{#1#2} \right \rangle} 
\newcommand{\Var}[1]{\mathrm{Var}[ #1 ]}
\newcommand{\zz}{\mathbb{Z}_2}

\DeclareBoldMathCommand\boldlangle{\langle} 
\DeclareBoldMathCommand\boldrangle{\rangle} 

\definecolor{deepblue}{rgb}{0,0,0.5}
\definecolor{deepred}{rgb}{0.6,0,0}
\definecolor{deepgreen}{rgb}{0,0.5,0}

\definecolor{keywords}{RGB}{255,0,90}
\definecolor{comments}{RGB}{150,0,255}
\definecolor{red}{RGB}{160,0,0}
\definecolor{green}{RGB}{0,150,0}
 \definecolor{darkblue}{rgb}{0.0, 0.0, 0.55}

\lstdefinestyle{pythonCode}{
language=Python,
basicstyle=\footnotesize,
keywordstyle=\bfseries\color{deepblue},
emphstyle=\bfseries, 
stringstyle=\color{deepgreen},
commentstyle=\color{comments},
numberstyle=\ttfamily\tiny,
numberblanklines=false,
frame=tb,                         
showstringspaces=false,            %
breaklines=true,
}

\lstdefinestyle{listXML}{
language=XML,
extendedchars=true, 
belowcaptionskip=5pt,
xleftmargin=1.8em,
xrightmargin=0.5em,
breaklines=true,
breakatwhitespace=true,
breakindent=0pt,
frame=tb,
breaklines=true,
emph={},
emphstyle=\color{red},
basicstyle=\small\ttfamily,
columns=fullflexible,
showstringspaces=false,
commentstyle=\color{gray}\upshape,
morestring=[b]",
morecomment=[s]{<?}{?>},
morecomment=[s][\color{comments}]{<!--}{-->}, 
keywordstyle=\normalfont\color{deepblue}, 
stringstyle=\normalfont\color{deepgreen}, 
tagstyle=\bfseries\color{deepblue}, 
morekeywords={xmlns,version,type,name,min,max,quantumnumber,matrixelement,change,source,target,default}
}

\newcommand{\eqnref}[1]{(\ref{#1})}

\long\def\beginmypgfpdfnamed#1#2\endmypgfpdfnamed{\includegraphics{#1}}

\graphicspath{{./pyfig/}}

\begin{document}

\begin{frontmatter}

\title{Matrix Product State applications for the ALPS project}

\author[eth-itp]{Michele Dolfi} \ead{dolfim@phys.ethz.ch}
\author[stationq]{Bela Bauer} 
\author[eth-qc]{Sebastian Keller} 
\author[eth-itp]{Alexandr Kosenkov} 
\author[unige-dpmc,bluebrain]{Timoth\'ee Ewart} 
\author[unige-dpmc]{Adrian Kantian} 
\author[unige-dpmc]{Thierry Giamarchi} 
\author[eth-itp]{Matthias Troyer} 

\address[eth-itp]{Theoretische Physik, ETH Zurich, 8093 Zurich, Switzerland}
\address[stationq]{Station Q, Microsoft Research, Santa Barbara, California 93106-6105, USA}
\address[eth-qc]{Laboratorium f\"ur Physikalische Chemie, ETH Zurich, 8093 Zurich, Switzerland}
\address[unige-dpmc]{DPMC-MaNEP, University of Geneva, 24 Quai Ernest-Ansermet, CH-1211 Geneva, Switzerland}
\address[bluebrain]{Blue Brain Project, Brain Mind Institute, EPFL, Switzerland}

\begin{abstract}
The density-matrix renormalization group method has become a standard computational approach to the low-energy physics as well as dynamics
of low-dimensional quantum systems. In this paper, we present a new set of applications, available
as part of the ALPS package, that provide an efficient and flexible implementation of these methods based on a matrix-product state (MPS)
representation.
Our applications implement, within the same framework, algorithms to variationally find the ground state and low-lying excited states as well as simulate the time evolution of arbitrary one-dimensional and two-dimensional models. Implementing the conservation of quantum numbers for generic Abelian symmetries, we achieve performance competitive with the best codes in the community.
Example results are provided for (i) a model of itinerant fermions in one dimension and (ii) a model of quantum magnetism.
\end{abstract}




\end{frontmatter}
{\bf PROGRAM SUMMARY}

\begin{small}
\noindent
{\em Program Title:} ALPS MPS \\
{\em Journal Reference:}                                      \\
{\em Catalogue identifier:}                                   \\
{\em Licensing provisions:} Use of `mps\_optim', 'mps\_tevol', 'mps\_meas' or 'mps\_overlap' requires citation of this paper. Use of any ALPS program requires citation of the ALPS \cite{bauer2011-alps} paper.\\
{\em Programming language:} \verb*#C++#, OpenMP for parallelization. \\
{\em Computer:} PC, HPC cluster \\ 
{\em Operating system:} Any, tested on Linux, Mac OS X and Windows\\ 
{\em RAM:} 100 MB - 100 GB.\\ 
{\em Number of processors used:} 1 - 24.
{\em Classification:} 7.7  \\ 
{\em External routines/libraries:}  ALPS \cite{bauer2011-alps, alps-web}, BLAS/LAPACK, HDF5.\\ 
{\em Nature of problem:} Solution of quantum many-body systems is generally a hard problem. The many-body Hilbert space grows exponentially with the system size which limits exact diagonalization results to only 20 − 40 spins, and the fermionic negative sign problem limits the Quantum Monte Carlo methods to a few special cases.\\
{\em Solution method:} The matrix product states ansatz provides a controllable truncation of the Hilbert space which makes it currently the method of choice to investigate low-dimensional systems in condensed matter physics.
Our implementation allows simulation of arbitrary one-dimensional and two-dimensional models and achieve performance competitive with the best codes in the community. We implement conservation of quantum numbers for generic Abelian symmetries.\\
{\em Running time:} 10s -- 8h per sweep.\\
\end{small}

\section{Introduction}
The density matrix renormalization group method (DMRG)~\cite{white1992} is currently the method of choice to investigate low-dimensional systems in condensed matter physics. Its applications range from the study of superconductivity, quantum magnetism and exotic phases of matter to simulation of quantum circuits.

On the theoretical side, a lot of progress has been made in understanding the success of this method on the basis of matrix product states (MPS)~\cite{ostlund1995}, which is the class of variational states underlying the DMRG method. This has facilitated the development of new algorithms, e.g. for real-time evolution~\cite{vidal2004,white2004,daley2004}, mixed-state (finite temperature) calculations~\cite{verstraete2004a,feiguin2005} as well as their combination~\cite{zwolak2004,barthel2009,pizorn2013}. These have been formulated directly in the language of MPS, and are now widely being applied and further developed.

Recently, there has also been a surge of applications of DMRG to two-dimensional systems~\cite{stoudenmire2012}. While it was realized early on that DMRG is exponentially hard for systems in more than one dimension~\cite{liang1994}, modern DMRG codes have become sufficiently efficient to combat this exponential scaling and reach system sizes where reliable conclusions about the thermodynamic limit of the 2d system become feasible~\cite{white2007,yan2011,bauer2012,depenbrock2012,jiang2012,bauer2014}.

In this paper we report the release into the ALPS project~\cite{albuquerque2007,bauer2011-alps,alps-web} of a set of MPS codes developed within the Swiss HP2C initiative~\cite{hp2c}, whose target was to develop new efficient and productive applications for current and future high performance computers. These applications have been designed to achieve the high performance required for the study of two-dimensional systems, and additionally to bind to the generic and well-known ALPS model and lattice libraries, which makes our MPS codes appealing for researchers in various fields. The new code has thus far been employed to study condensed matter physics problems such as ordering in an SU(3) Heisenberg model~\cite{bauer2012}, the analysis of critical points for a system of supersymmetric fermions~\cite{bauer2013}, the analysis of interacting mesoscopic structures~\cite{cheng2013}, spin liquid phases in a Kagome-lattice spin model~\cite{bauer2014}, and recent method developments such as the applications of DMRG as impurity solver within DMFT~\cite{shinaoka2014} and the development a multigrid DMRG scheme~\cite{dolfi2012},

\begin{table}[h]
\begin{center}
\caption{Runtime of our code compared with other state-of-the-art condensed matter DMRG codes. The benchmark is a run of 6 sweeps with a maximum bond dimension of $M=200$ that increases linearly with every sweep for an antiferromagnetic Heisenberg chain with $L=100$ sites. All codes are compiled with GCC 4.8.1, optimization option \textrm{-O3} and linked against the sequential version of Intel MKL v11.1.1. Runtime is measured on a 2.2 GHz Intel Ivy Bridge EP E5-2660v2 processor. While we have done our best to set the parameters for all codes such as to ensure comparable executions, these timings should be understood as examples and results may depend significantly on the parameters for each code.}
\vspace{1em}
\begin{tabular}{p{.6\linewidth}c}
\toprule
{\bf Code}   & {\bf Time}  \\ \midrule
ALPS MPS (our new code)  & 16 sec \\
ALPS DMRG~\cite{albuquerque2007,bauer2011-alps} & 73 sec \\
ITensor~\cite{itensor} & 24 sec \\
OSMPS~\cite{openmps} & 40 sec \\
\bottomrule
\end{tabular}
\label{tab:benchmarks}
\end{center}
\end{table}

A number of open-source implementations of DMRG are currently available~\cite{bauer2011-alps,openmps,itensor,alvarez2009,block-web,wouters2014}.
With this release, we add a set of applications that, while implementing a generic framework amenable to many extensions, offers
performance on par with the best available codes (see Table~\ref{tab:benchmarks} for a timing example). Furthermore, it is seamlessly integrated with the widely used ALPS libraries.

The outline of this paper is as follows:
In Section~\ref{sec:algo-implement} we first focus on the characteristics and conventions of the new MPS code. In Section~\ref{sec:code-example} we explain the main application parameters and show the results of a tutorial. For the theoretical background on MPS and DMRG we point the reader to the review~\cite{schollwock2011} and references therein.

\section{Algorithms and implementation}
\label{sec:algo-implement}
The matrix product state (MPS) is a variational ansatz wave function which for a chain of $N$ sites reads
\begin{equation}
  \label{eq:mps}
  \ket{\psi} = \sum_{\gv \sigma} A_1^{\sigma_1} A_2^{\sigma_2} \cdots A_N^{\sigma_N} \ket{\gv \sigma},
\end{equation}
where the exponentially growing number of parameters of the full quantum state is efficiently truncated to a polynomial number $\propto LM^2$ contained in the $M\times M$ matrices $A_i^{\sigma_i}$. This introduces an approximation that can be controlled by the matrix size $M$.

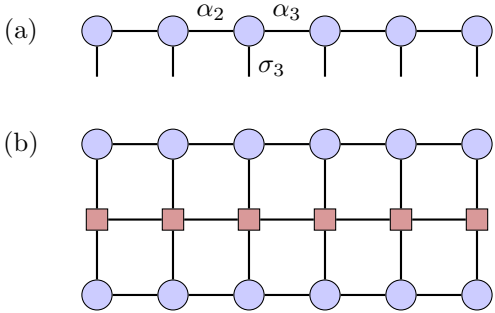
\begin{figure}
  \centering
  \beginmypgfpdfnamed{fig_mps}

\begin{tikzpicture}
\def\pos{0}
\def\ppos{-1.5}
\def\pppos{-4}

\node at (0,\pos) {(a)};

\foreach \i in {1,...,6} {
	\node[spin] (M1 \i) at (\i,\pos) {};
	\draw[thick] (M1 \i.south) -- +(0,-.4);
}
\foreach \i [evaluate = \i as \j using int(\i+1)] in {1,...,5} {
	\draw[thick] (M1 \i) -- (M1 \j);
}

\node at (2.5, .25) {$\alpha_2$};
\node at (3.5, .25) {$\alpha_3$};
\node at (3.3, -.5) {$\sigma_3$};

\node at (0,\ppos) {(b)};

\foreach \i in {1,...,6} {
	\node[spin]  (M2a \i) at (\i,\ppos) {};
	\node[hamil] (H2 \i) at (\i,\ppos-1) {};
	\node[spin]  (M2b \i) at (\i,\ppos-2) {};
  
	\draw[thick] (M2a \i.south) -- (H2 \i.north);
	\draw[thick] (H2 \i.south)  -- (M2b \i.north);
}
\foreach \i [evaluate = \i as \j using int(\i+1)] in {1,...,5} {
	\draw[thick] (M2a \i) -- (M2a \j);
	\draw[thick] (H2 \i)  -- (H2 \j);
	\draw[thick] (M2b \i) -- (M2b \j);
}

\end{tikzpicture}
\endmypgfpdfnamed
  \caption{
  (a) Graphical representation of the MPS. Vertices (blue dots) represent the trivalent tensor on the $i$'th site, $(A^{\sigma_i}_i)_{\alpha_{i-1}\alpha_i}$, where edges indicate indices. Closed edges, such as the horizontal lines indicate the indices $\alpha_{i-1}$, $\alpha_i$ that are being contracted; vertical, open edges indicate the physical indices $\sigma_i$.
  (b) Contraction of the observable $\langle \psi | \hat O | \psi \rangle$, where $\hat O$ is represented as a matrix-product operator (red squares). 
  }
  \label{fig:mps graphics}
\end{figure}

Matrix-product states are an example of a tensor network state. Such states are compactly represented
graphically, as shown in Fig.~\ref{fig:mps graphics}. A tensor network is represented as
a graph where vertices represent tensors, and edges represent indices that are contracted. Thus, in the upper
panel of Fig.~\ref{fig:mps graphics}, the blue circles represent the tensors $\left( A_i^{\sigma_i} \right)_{\alpha_{i-1}\alpha_i}$
of Eqn.~\eqnref{eq:mps}; the edges connecting the tensors represent the indices that are summed over
to yield matrix multiplication; and the open vertices pointing down represent the physical indices $\sigma_i$.
The lower panel of the figure shows a closed tensor network, i.e. without any open legs, that thus corresponds
to a scalar. The contraction shown yields the calculation of an expectation value for an operator represented
by a matrix-product operator (see Section~\ref{sec:mpo}).



It has been shown that ground states of gapped Hamiltonians in one dimension are accurately represented by a matrix-product state with a matrix size that grows only polynomially in the system size~\cite{hastings2007,schuch2008-1}; if only local properties are required, one can obtain even better scaling~\cite{verstraete2006-2}.
Intuitively, this result rests on the fact that an area law holds for these systems~\cite{eisert2010colloquium}, i.e. the bipartite entanglement entropy in these states is a function only of the area of the cut between two regions and not the total system size. In one dimension, this leads to constant bipartite entanglement entropy for gapped states.

The MPS ansatz is inherently a one-dimensional open chain. To simulate systems on periodic systems, ladders or even two-dimensional systems, a mapping from these lattices to a chain has to be chosen. This will lead to a Hamiltonian with longer-ranged interactions on the chain, where the range of interactions is related to the width of the ladder or 2d system. More importantly, the entanglement structure of the state is affected. The area of a cut bipartitioning the system into two halves will generally grow with the linear size $L$ of the system. Since the bond dimension required to capture entropy $S$ in an MPS is exponential, $M \sim e^S$, this implies an exponential cost for MPS simulations for two-dimensional systems.
The specific mapping from the 2d system to the chain therefore may have great impact on the accuracy and performance of simulations, and should therefore be chosen very carefully to minimize entanglement across each cut.



\subsection{Representation of Matrix Product Operators}
\label{sec:mpo}

To introduce the notion of matrix product operators, it is useful to consider a 1d quantum lattice system with the same Hilbert space $\mathcal{H}$ on each site. Let $\lbrace \hat{O}^\beta_k \rbrace$ be the set of Hermitian operators that can act on the $k$'th site; as an example, these are identity matrix and the three Pauli matrices, $\lbrace \mathbb{I}, \sigma^x, \sigma^y, \sigma^z \rbrace$ for the case of $S=1/2$ spins. We can then write any Hermitian operator $\hat{Q}$ acting on the entire lattice as
\begin{equation}
\hat{Q} = \sum c(\beta_1, \ldots)\  \hat{O}^{\beta_1}_1 \otimes \hat{O}^{\beta_2}_2 \otimes \ldots.
\end{equation}
This makes use of the fact that the operators themselves form a Hilbert space, analogous to quantum state. The above expression suggests that the coefficients $c(\alpha_1, \ldots)$ can be decomposed into a matrix product, analogous to Eqn.~\ref{eq:mps}. This representation is referred to as matrix-product operator (MPO). Analogous to a matrix-product state, such a matrix-product operator has an associated matrix size. Local Hamiltonians can generally be represented as an MPO of low bond dimension; even for long-range Hamiltonians, an efficient MPO description is often possible.

Consider now writing a local Hamiltonian, say for free fermions,
\begin{equation}
  \label{eq:free fermions}
  \hat H = -\sum_{\langle ij \rangle} \left( \hat c^\dagger(i)\, \hat c(j) + \text{h.c.} \right) \notag
\end{equation}
in this form. For simplicity, let's consider this model on three sites but with
additional next-nearest neighbor hopping:
\begin{equation}
  \hat H = -\hat c_1^\dagger \hat c_2 + \hat c_1 \hat c_2^\dagger - \hat c_2^\dagger \hat c_3 + \hat c_2 \hat c_3^\dagger -
   \hat c_1^\dagger \hat c_3 + \hat c_1 \hat c_3^\dagger
\end{equation}
The most naive way to perform this decomposition is to assign an independent auxiliary index to each term in each of the sums:
\begin{align}
H &= \sum_{k,k'} A_{1k} B_{kk'} C_{k'1} \\
A_{1k} &= \left( -\hat c_1^\dagger, \hat c_1, -\mathbb{\hat I}, \mathbb{\hat I}, -\hat c_1^\dagger, \hat c_1 \right)_k \\
B_{kk'} &= \left( \hat c_2, \hat c_2^\dagger, \hat c_2^\dagger, \hat c_2, \mathbb{\hat I}, \mathbb{\hat I} \right)_k \delta_{k k'} \\
C_{k1} &= \left( \mathbb{\hat I}, \mathbb{\hat I}, \hat c_3, \hat c_3^\dagger, \hat c_3, \hat c_3^\dagger \right)_k
\end{align}
This leads to an MPO of bond dimension 6, which is very inefficient and redundant. A more
efficient version is the following:
\begin{align}
\tilde{A}_{1k} &= \left( \mathbb{\hat I}, \hat c_1^\dagger, \hat c_1 \right)_k \\
\tilde{B}_{k k'} &= \left(
    \begin{array}{ccc}
    0 &\hat c_2 &-\hat c_2^\dagger \\
    -\hat c_2 &0 &-\mathbb{\hat I} \\
    \hat c_2^\dagger &\mathbb{\hat I} &0
    \end{array} \right)_{kk'} \\
\tilde{C}_{k1} &= \left( \mathbb{\hat I}, \hat c_3^\dagger, \hat c_3 \right)_k.
\end{align}
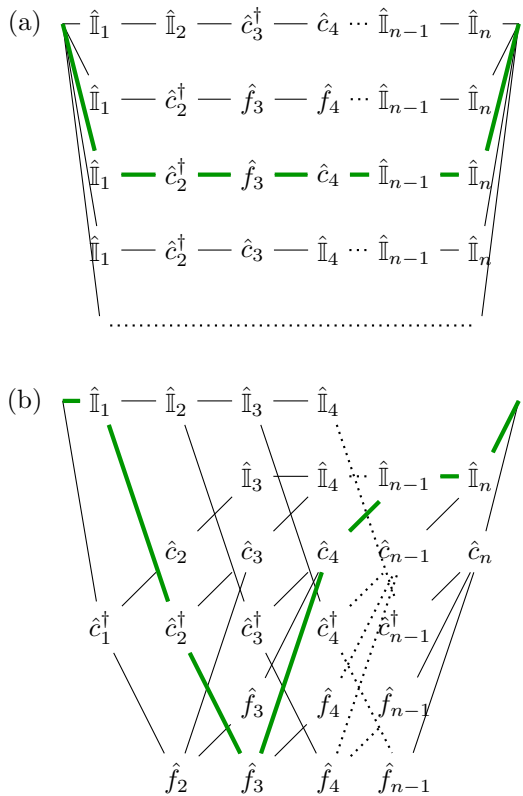
\begin{figure}
  \centering
\beginmypgfpdfnamed{fig_mpo}

\tikzset{mpo_highlight/.style={ultra thick,draw=green!100}}

\begin{tikzpicture}
\def\pos{0}

\node at (0,\pos) {(a)};

\newcommand\OpLine[2]{
  
  \foreach[count=\i] [evaluate = \i as \j using int(\i-1)] \argtext in {#2} {
    \node (Op#1_\i) at (\i,#1) {\argtext};
            \ifnum\i=1
            \else
              \draw (Op#1_\i) -- (Op#1_\j);%
            \fi%
  }
}
\OpLine{0}{  {$\mathbb{\hat I}_1$}, {$\mathbb{\hat I}_2$}, {$\hat c^\dagger_3$}, {$\hat c_4$} }
\OpLine{-1}{ {$\mathbb{\hat I}_1$}, {$\hat c^\dagger_2$},  {$\hat f_3$},         {$\hat f_4$} }
\OpLine{-2}{ {$\mathbb{\hat I}_1$}, {$\hat c^\dagger_2$},  {$\hat f_3$},         {$\hat c_4$} }
\OpLine{-3}{ {$\mathbb{\hat I}_1$}, {$\hat c^\dagger_2$},  {$\hat c_3$},         {$\mathbb{\hat I}_4$} }

\draw (0.5,0) -- (Op0_1);
\draw (0.5,0) -- (Op-1_1);
\draw (0.5,0) -- (Op-2_1);
\draw (0.5,0) -- (Op-3_1);

\node (Op0_5)  at (5, 0)  {$\mathbb{\hat I}_{n-1}$};
\node (Op-1_5) at (5, -1) {$\mathbb{\hat I}_{n-1}$};
\node (Op-2_5) at (5, -2) {$\mathbb{\hat I}_{n-1}$};
\node (Op-3_5) at (5, -3) {$\mathbb{\hat I}_{n-1}$};

\draw[thick, dotted] (Op0_4)  -- (Op0_5);
\draw[thick, dotted] (Op-1_4) -- (Op-1_5);
\draw[thick, dotted] (Op-2_4) -- (Op-2_5);
\draw[thick, dotted] (Op-3_4) -- (Op-3_5);

\node (Op0_6)  at (6, 0)  {$\mathbb{\hat I}_n$};
\node (Op-1_6) at (6, -1) {$\mathbb{\hat I}_n$};
\node (Op-2_6) at (6, -2) {$\mathbb{\hat I}_n$};
\node (Op-3_6) at (6, -3) {$\mathbb{\hat I}_n$};

\draw (Op0_5)  -- (Op0_6);
\draw (Op-1_5) -- (Op-1_6);
\draw (Op-2_5) -- (Op-2_6);
\draw (Op-3_5) -- (Op-3_6);

\draw (Op0_6) --  (6.5, 0);
\draw (Op-1_6) -- (6.5, 0);
\draw (Op-2_6) -- (6.5, 0);
\draw (Op-3_6) -- (6.5, 0);

\node (Op-4_1) at (1,-4) {};
\node (Op-4_6) at (6,-4) {};
\draw (0.5,0) -- (Op-4_1);
\draw[thick, dotted] (Op-4_1) -- (Op-4_6);
\draw (Op-4_6) -- (6.5, 0);

\draw[mpo_highlight] (0.5,0) -- (Op-2_1) -- (Op-2_2) -- (Op-2_3) -- (Op-2_4) -- (Op-2_5) -- (Op-2_6) -- (6.5, 0);

\def\pos{-5}
\node at (0,\pos) {(b)};

\node (OO11) at (1,\pos-0) {$\mathbb{\hat I}_1$};
\node (OO14) at (1,\pos-3) {$\hat c^\dagger_1$};

\node (OO21) at (2,\pos-0) {$\mathbb{\hat I}_2$};
\node (OO23) at (2,\pos-2) {$\hat c_2$};
\node (OO24) at (2,\pos-3) {$\hat c^\dagger_2$};
\node (OO26) at (2,\pos-5) {$\hat f_2$};

\foreach \i in {3,...,4} {
  \node (OO\i1) at (\i,\pos-0) {$\mathbb{\hat I}_\i$};
  \node (OO\i2) at (\i,\pos-1) {$\mathbb{\hat I}_\i$};
  \node (OO\i3) at (\i,\pos-2) {$\hat c_\i$};
  \node (OO\i4) at (\i,\pos-3) {$\hat c^\dagger_\i$};
  \node (OO\i5) at (\i,\pos-4) {$\hat f_\i$};
  \node (OO\i6) at (\i,\pos-5) {$\hat f_\i$};
}

\node (OO52) at (5,\pos-1) {$\mathbb{\hat I}_{n-1}$};
\node (OO53) at (5,\pos-2) {$\hat c_{n-1}$};
\node (OO54) at (5,\pos-3) {$\hat c^\dagger_{n-1}$};
\node (OO55) at (5,\pos-4) {$\hat f_{n-1}$};
\node (OO56) at (5,\pos-5) {$\hat f_{n-1}$};

\node (OO62) at (6,\pos-1) {$\mathbb{\hat I}_n$};
\node (OO63) at (6,\pos-2) {$\hat c_n$};

\draw (0.5,\pos-0) -- (OO11);
\draw (0.5,\pos-0) -- (OO14);

\draw (OO11) -- (OO21);
\draw (OO11) -- (OO24);
\draw (OO14) -- (OO23);
\draw (OO14) -- (OO26);

\draw (OO21) -- (OO31);
\draw (OO21) -- (OO34);
\draw (OO23) -- (OO32);
\draw (OO24) -- (OO33);
\draw (OO24) -- (OO36);
\draw (OO26) -- (OO33);
\draw (OO26) -- (OO35);

\draw (OO31) -- (OO41);
\draw (OO31) -- (OO44);
\draw (OO32) -- (OO42);
\draw (OO33) -- (OO42);
\draw (OO34) -- (OO43);
\draw (OO34) -- (OO46);
\draw (OO35) -- (OO43);
\draw (OO36) -- (OO43);
\draw (OO36) -- (OO45);

\draw[thick, dotted] (OO41) -- (OO54);
\draw[thick, dotted] (OO42) -- (OO52);
\draw[thick, dotted] (OO43) -- (OO52);
\draw[thick, dotted] (OO44) -- (OO53);
\draw[thick, dotted] (OO44) -- (OO56);
\draw[thick, dotted] (OO45) -- (OO53);
\draw[thick, dotted] (OO46) -- (OO53);
\draw[thick, dotted] (OO46) -- (OO55);

\draw (OO52) -- (OO62);
\draw (OO53) -- (OO62);
\draw (OO54) -- (OO63);
\draw (OO55) -- (OO63);
\draw (OO56) -- (OO63);

\draw (OO62) -- (6.5,\pos-0);
\draw (OO63) -- (6.5,\pos-0);

\draw[mpo_highlight] (0.5,\pos-0) -- (OO11) -- (OO24) -- (OO36) -- (OO43) -- (OO52) -- (OO62) -- (6.5, \pos-0);

\end{tikzpicture}
\endmypgfpdfnamed
  \caption{Internal structure of the MPO representing the the first part $\hat c^\dagger(i)\, \hat c(j)$ of the Hamiltonian \eqref{eq:free fermions} on a $3 \times 3$ square lattice.
  (a) trivial graph with an auxiliary bond for each term.
  (b) compressed version of the same graph where common branches are computed only once.
For comparison, we highlight the term $\hat c^\dagger(2)\, \hat c(4)$ in both panels.}
  \label{fig:mpo tree}
\end{figure}
This MPO implements the same Hamiltonian, but has only bond dimension 3 by making use of the fact
that the operators $\hat c_1^\dagger$, $\hat c_1$ as well as $\hat c_3^\dagger$, $\hat c_3$ occur both in the nearest- as
well as next-nearest neighbor hopping term. Note that in the notation above, we have taken
the elements of the matrices to be fermionic operators themselves. The antisymmetry of the central matrix
in this case is a special property of the 3-site ring and reflects the anticommutation of fermionic operators. In the
actual implementation, fermionic statistics is implemented via a Jordan-Wigner transformation.

Our code implements an optimized MPO network that merges common branches, as illustrated in the above
example, thus reducing the matrix size of the MPO substantially. This is sketched in the lower panel of
Figure~\ref{fig:mpo tree} (for a more extensive explanation of similar notation, see Ref.~\cite{crosswhite2008}).
Note that in principle, the same compression techniques that are available
for reducing the matrix size of MPSs can also be applied to MPOs; however, in some cases constructing
the MPO in a naive way and then compressing is prohibitively time-consuming. Therefore, the code is
written to perform many optimizations when the MPO is first constructed.

\subsubsection{Fermionic operators}
Special treatment is needed to account for the anti-commutation relations $\{\hat c(i), \hat c^\dagger(j)\} = \delta_{ij}$ characteristic of fermions. All fermionic operators have to be converted into a string of bosonic operators according to the Jordan-Wigner transformation~\cite{Jordan1928}:
\begin{equation}
  \hat c(i) = \left(\prod_{k<i} \hat f(k)\right) \hat b(i),
\end{equation}
where $\hat f(k)$ is the fermion sign operator, a diagonal matrix with a $-1$ entry for each local basis state with an odd number of fermions, and $1$ for each local basis state with an even (including 0) number of fermions, and $\hat b(i)$ are hard-core boson operators satisfying the commutation relation $[\hat b(i), \hat b^\dagger(j)] = \delta_{ij}$ and $(\hat b(i))^2 = 0$. The fermion sign matrix $\hat f(k)$ can be written as $\hat f(k) = 1- 2 \hat b^\dagger(k) \hat b(k)$. To represent spinful fermions, two flavors of bosons $\hat b_1(k)$, $\hat b_2(k)$ have to be introduced and a normal-ordering of operators has to be chosen; in the ALPS libraries, the ordering is chosen as $\hat b_1(1) \hat b_2(1) \hat b_1(2) \hat b_2(2) ...$. The Jordan-Wigner string must then include fermion sign operators for both flavors of bosons, i.e. two per physical site.

Noting that $\hat f^2 = \mathbb{I}$ and $[\hat f(i), \hat b^{(\dagger)}(j)] = 0$ for $i \neq j$, we find that for a fermionic hopping term the transformation simplifies to just a sequence of $\hat f(k)$ matrices between the two sites $i$ and $j$:
\begin{equation}
  \hat c(i) \hat c^\dagger(j) = \hat b(i) \left(\prod_{i\le k<j} \hat f(k)\right) \hat b^\dagger(j). 
\end{equation}
A similar simplification will apply to other types of fermionic bond terms; bond terms where each term is diagonal, such as density-density interactions, do not require Jordan-Wigner transformation.

\subsection{Variational optimization of the MPS}
Given the variational nature of the MPS ansatz, the ground state of a model described by the Hamiltonian $\hat H$ is obtained by an optimization process of all variational parameters looking for the energy minimum.

The most common way to optimize an MPS, as already described by S. White in '92~\cite{white1992} is to optimize one or two tensors at the time, keeping all the others constant. For the case of the single site optimization the equation to solve is
\begin{equation}
\pd{}{A_i^*} \left(  \langle \psi | \hat H | \psi \rangle - \lambda \left[\braket{\psi}{\psi} - 1\right] \right) = 0,
\end{equation}
which maps to solving for the lowest eigenvector of the generalized eigenvalue problem $\mathcal{M} A_i - \lambda \mathcal{N} A_i = 0$, which is solved iteratively with the Jacobi-Davison algorithm provided by the IETL package in ALPS. Note that in the iterative solver the \emph{matrix} $\mathcal{M}$ is never computed, but we efficiently contract all tensor legs on the fly in order to keep a complexity $\mathcal{O}(M^3)$.

Convergence is reached after sweeping a few times through the system. One \emph{sweep} is here defined as optimizing twice all tensors in the chain, i.e. moving from left to right and then from right to left.

In our \verb|mps_optim| application we implement the so called \emph{single-site} and \emph{two-site} optimizations. The first one optimizes one tensor at the time, whereas the latter one fuses two sites together to optimize the bond between them. The user is free to choose between them by changing the \verb|optimization| input parameter to \verb|singlesite| or \verb|twosite|, respectively. It is strongly advised to always play around with both, in order to find what fits better the system to be solved.

In the case of the single site optimization we implement the procedure proposed in Ref.~\cite{white2005}. Here, the two shortcomings of the single-site optimization -- the possibility of trapping on local minima and the inability to change the bond dimension dynamically -- are remedied by enlarging the reduced density matrix after each optimization using
\begin{equation*}
\tilde \rho^{A} = \mathrm{Tr}_B |\psi\rangle \langle \psi |  + \alpha \sum_{b_l} \mathrm{Tr}_B \hat H^A_{b_l} |\psi \rangle \langle \psi| \hat H^A_{b_l}.
\end{equation*}
Assuming a partition of the system into a sub-block $A$ and sub-block $B$ such that the Hamiltonian is rewritten as $\hat H = \sum_{b_l} \hat H^A_{b_l} \hat H^B_{b_l}$, the first term in $\tilde \rho^{A}$ is the usual reduced density matrix of the subsystem $A$, whereas the second term introduces some reshuffling of the truncated states that protects against trapping in the optimization into local minima. The parameter $\alpha$ is a small number -- typically $10^{-4}$ -- which is slowly taken to zero during sweeping. In our implementation we allow three values for it {\tt alpha\_initial}, {\tt alpha\_main} and {\tt alpha\_final} (see the parameters in Table~\ref{tab:optim params} for further details).

\subsubsection{Excited states}
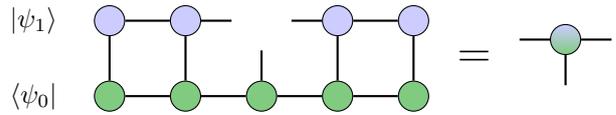
\begin{figure}
  \centering
  \beginmypgfpdfnamed{fig_ortho_mps}

\begin{tikzpicture}
\def\pos{0}
\def\ppos{-1.5}
\def\pppos{-4}

\node at (0,\pos) {$|\psi_1 \rangle$};
\node at (0,\pos-1) {$\langle \psi_0 |$};

\foreach \i in {1,...,2} {
	\node[spin]  (Ma \i) at (\i,\pos) {};
	\node[spin,fill=green!50] (Mb \i) at (\i,\pos-1) {};
  
	\draw[thick] (Ma \i.south) -- (Mb \i.north);
}
\foreach \i [evaluate = \i as \j using int(\i+1)] in {1,...,1} {
	\draw[thick] (Ma \i) -- (Ma \j);
	\draw[thick] (Mb \i) -- (Mb \j);
}

\foreach \i in {4,...,5} {
	\node[spin]  (Ma \i) at (\i,\pos) {};
	\node[spin,fill=green!50] (Mb \i) at (\i,\pos-1) {};
  
	\draw[thick] (Ma \i.south) -- (Mb \i.north);
}
\foreach \i [evaluate = \i as \j using int(\i+1)] in {4,...,4} {
	\draw[thick] (Ma \i) -- (Ma \j);
	\draw[thick] (Mb \i) -- (Mb \j);
}

\node[spin,fill=green!50] (Mb 3) at (3,\pos-1) {};
\draw[thick] (Mb 2) -- (Mb 3);
\draw[thick] (Mb 3) -- (Mb 4);



\draw[thick] (Ma 2.east) -- +(0.4,0);
\draw[thick] (Mb 3.north) -- +(0,0.4);
\draw[thick] (Ma 4.west) -- +(-0.4,0);

\node at (5.8,\pos-0.5) {\LARGE $=$};

\node[spin,  top color=blue!20, bottom color=green!50] (Mc) at (7,\pos-0.25) {};
\draw[thick] (Mc.east) -- +(0.4,0);
\draw[thick] (Mc.west) -- +(-0.4,0);
\draw[thick] (Mc.south) -- +(0,-0.4);

\end{tikzpicture}
\endmypgfpdfnamed
  \caption{Construction of the environment tensor used as a local orthogonality constraint. The orthogonality constraint shown in the left network corresponds to a \emph{mixed} MPS tensor which is used to evaluate $\langle \psi_0 | \psi_1\rangle$ in the local optimization.}
  \label{fig:ortho mps}
\end{figure}

In traditional DMRG approaches, the calculation of excited states is implemented through multi-state targetting, where the reduced density matrix after each iteration is truncated to describe a number of low-lying states simultaneously. While this approach has been shown to work reasonably, a more controlled approach is to sequentially find low-lying states by first finding the ground state $|\psi_0\rangle$, then finding the lowest-energy state $|\psi_1\rangle$ that is orthogonal to the ground state, $\langle \psi_0 | \psi_1 \rangle = 0$, etc. This approach was first described in Ref.~\cite{mcculloch2007}.

In the MPS context, the overlap of two states can be efficiently calculated; furthermore, when optimizing a local tensor, the orthogonality constraint can be expressed as a local constraint of orthogonalizing the MPS tensor against the environment tensor shown in Fig.~\ref{fig:ortho mps}. For higher excited states, several such constraints need to be taken into account. In an iterative eigensolver, such as the Jacobi-Davidson method used here, these constraints can be taken into account by projecting newly generated Krylov vectors (or, in the case of Jacobi-Davidson, search vectors) into the orthogonal complement of the space spanned by the constraints. When sweeping through the state, the environment of Fig.~\ref{fig:ortho mps} can be updated at very low computational cost.

\subsection{Time evolution}
To simulate time evolution, we follow the approach of Refs.~\cite{vidal2003,daley2004,white2004}.
The key insight to performing time evolution using MPS is that instead of time evolution, one can also variationally optimize the difference between an MPS $|\psi_1\rangle$ and an MPS with a unitary operator applied, $U |\psi_0\rangle$; that is, the optimization problem
\begin{equation}
  \min_{|\psi_1\rangle} \big\| |\psi_1\rangle - U |\psi_0\rangle \big\|
\end{equation}
can be solved efficiently over normalized MPS. Here, the operator $U$ can either act locally on a few sites, or be a sum of terms expressed as an MPO. Another key principle is that the evolution under a local Hamiltonian, $\hat H = \sum_n \hat h_n$, can be expressed as a product over local unitaries using a Trotter-Suzuki decomposition~\cite{trotter1959,suzuki1976}. Here, the time evolution operator $\exp(-i \hat H t)$ for a small time step $\Delta t$ is decomposed into multiple products of the non-communing terms in the Hamiltonian. To first order the Trotter-Suzuki decomposition is
\begin{equation}
  \label{eq:trotter 1st}
  \exp(-i \hat H t) = \prod_{k=1}^{t/\Delta t} \prod_n e^{-i \hat h_n \Delta t} + \mathcal{O}(\Delta t^2),
\end{equation}
where $\hat h_n$ can themselves be sums of terms that do not overlap and hence commute.
In the case of a simple nearest-neighbor Hamiltonian on a chain, $\hat h_1$ and $\hat h_2$ are the terms acting on even and odd bonds, respectively. In general better accuracy is obtained by second and fourth orders which are implemented in the \verb|mps_evolve| application (see \ref{app:trotter}).

Naively, the application of the unitary to an MPS increases the bond dimension to $M' = M\cdot D_h$, where $D_h > 1$ depends on the structure of the Hamiltonian; for details, see below.
Therefore, after each contraction, the MPS matrices are truncated to their original size either with an singular value decomposition, or with a variational approximation (see Ref.~\cite{schollwock2011}). Our application implements two types of tensor contractions for time evolution, \emph{nearest neighbors} and \emph{mpo}.

\paragraph{Nearest neighbors} Nearest neighbors time evolution algorithm is an optimized version used when the Hamiltonian contains only nearest neighbors bond term (in the case of ladders or 2d systems after having been mapped it to a Hamiltonian on a chain). The bond gates are contracted with the MPS tensors and a truncation via singular value decomposition is applied on the fly, following the TEBD method described in Ref.~\cite{vidal2003}.

\paragraph{MPO} MPO time evolution is a more generic algorithm where each $\exp(-i \hat h_n \Delta t)$ is transformed into an MPO, which allows to encode bond terms between non-neighboring sites. After multiplying the MPS wave function with the MPO, this is truncated to its original size $M$ with a variational compression. Special care has to be taken for fermionic long-range operators, this procedure is explained in \ref{app:fermion exponential}.

\subsection{Efficient tensor storage with abelian symmetries}
Symmetries manifest themselves in many physical models, i.e. there are global operators $\hat G$ that commute with the system Hamiltonian, $[\hat H, \hat G] = 0$. In such a case, due to the fact that these operators can be simultaneously diagonalized, $\hat H$ becomes block-diagonal in a basis of eigenstates of $\hat G$. The blocks of the Hamiltonian are labelled by {\it quantum numbers}, corresponding to possible eigenvalues of $\hat G$. It is often computationally favorable to diagonalize the Hamiltonian separately within each quantum number sector.

In the case of DMRG, this is particularly easy when the symmetry has local generators, such that the reduced density matrices for any part of the system can also be block-diagonalized. Since the bond indices of the matrix product state essentially enumerate eigenstates of reduced density matrices, these can be associated with quantum number sectors. Each matrix of the MPS then obeys the symmetry locally, i.e. there are local conservation laws, and the matrices also become block-diagonal.


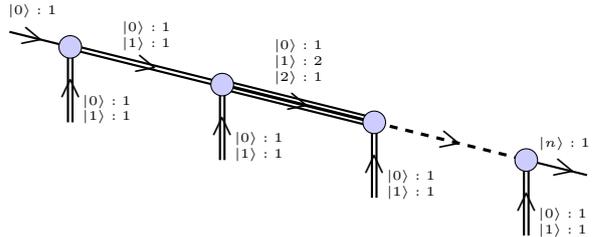
\begin{figure}
  \centering
  \beginmypgfpdfnamed{fig_mps_symemtries2}

\begin{tikzpicture}
\def\pos{0}
\def\ppos{-1.5}
\def\pppos{-4}

\node (M1) at (0,  0  ) {};
\node (M2) at (2, -0.5) {};
\node (M3) at (4, -1  ) {};
\node (Mn) at (6, -1.5) {};

\draw[thick] ($(M1)+(-0.03,0)$) -- +(0,-1);
\draw[thick] ($(M1)+(+0.03,0)$) -- +(0,-1) node [midway, below right, align=left,font=\tiny] {$| 0 \rangle : 1$ \\ $| 1 \rangle : 1$};
\node[rotate=90, font=\bfseries] at ($(M1)+(0,-0.55)$) {\textgreater};

\draw[thick] ($(M2)+(-0.03,0)$) -- +(0,-1);
\draw[thick] ($(M2)+(+0.03,0)$) -- +(0,-1) node [midway, below right, align=left,font=\tiny] {$| 0 \rangle : 1$ \\ $| 1 \rangle : 1$};
\node[rotate=90, font=\bfseries] at ($(M2)+(0,-0.55)$) {\textgreater};

\draw[thick] ($(M3)+(-0.03,0)$) -- +(0,-1);
\draw[thick] ($(M3)+(+0.03,0)$) -- +(0,-1) node [midway, below right, align=left,font=\tiny] {$| 0 \rangle : 1$ \\ $| 1 \rangle : 1$};
\node[rotate=90, font=\bfseries] at ($(M3)+(0,-0.55)$) {\textgreater};

\draw[thick] ($(Mn)+(-0.03,0)$) -- +(0,-1);
\draw[thick] ($(Mn)+(+0.03,0)$) -- +(0,-1) node [midway, below right, align=left,font=\tiny] {$| 0 \rangle : 1$ \\ $| 1 \rangle : 1$};
\node[rotate=90, font=\bfseries] at ($(Mn)+(0,-0.55)$) {\textgreater};

\draw[thick] ($(M1)$) -- +(-.8,+.2);
\node[rotate=-14, font=\bfseries] at (-.5,0.125) {\textgreater};
\node at (-.5,0.475) {\tiny $| 0 \rangle : 1$};

\draw[thick] ($(M1)+(0,0.03)$) -- ($(M2)+(0,0.03)$);
\draw[thick] ($(M1)+(0,-0.03)$) -- ($(M2)+(0,-0.03)$);
\node[rotate=-14, font=\bfseries] at (1,-0.25) {\textgreater};
\node[font=\tiny, align=left] at (1,0.15) { $| 0 \rangle : 1$ \\ $| 1 \rangle : 1$};

\draw[thick] ($(M2)+(0,0.06)$) -- ($(M3)+(0,0.06)$);
\draw[very thick] ($(M2)$) -- ($(M3)$);
\draw[thick] ($(M2)+(0,-0.06)$) -- ($(M3)+(0,-0.06)$);
\node[rotate=-14, font=\bfseries] at (3,-0.75) {\textgreater};
\node[font=\tiny, align=left] at (3,-0.2) { $| 0 \rangle : 1$ \\ $| 1 \rangle : 2$ \\ $| 2 \rangle : 1$};

\draw[very thick, dashed] ($(M3)$) -- ($(Mn)$);
\node[rotate=-14, font=\bfseries] at (5,-1.25) {\textgreater};

\draw[thick] ($(Mn)$) -- +(+.8,-.2);
\node[rotate=-14, font=\bfseries] at (6.5,-1.625) {\textgreater};
\node at (6.5,-1.275) {\tiny $| n \rangle : 1$};

\node[spin small] (M1 tens) at (0,  0  ) {};
\node[spin small] (M2 tens) at (2, -0.5) {};
\node[spin small] (M3 tens) at (4, -1  ) {};
\node[spin small] (Mn tens) at (6, -1.5) {};

\end{tikzpicture}
\endmypgfpdfnamed
  \caption{Labels in an MPS with conserved particles number. In the pair $(|k\rangle, s)$ we describe a charge block with label $|k\rangle$ ($k$ particles) of size $s$. The MPS always starts with the trivial block $(|0\rangle, 1)$ and ends with the total charge sector that the user wants to target, $n$ particles in this example.}
  \label{fig:mps symmetries}
\end{figure}

For each bond of the tensor network, a gauge choice has to be made fixing the direction of the bond. When considering, for example, particle number conservation, this is a choice of which bonds are considered going into and which coming out of the node of the network~\cite{bauer2011}.
The choice we make in our implementation is shown in Fig.~\ref{fig:mps symmetries}.
The MPS always starts with a trivial index on the left (e.g. only one sector with zero particles), new states are then added on every site from the local Hilbert space $\sigma_i$, and eventually the last tensor is fixed to the total quantum number sector chosen by the user.

Given our convention, locally the conservation of quantum numbers reads
\begin{equation*}
  \sigma_i \circ \alpha_i = \alpha_{i+1} \quad \text{and} \quad \alpha_i = (-\sigma_i) \circ \alpha_{i+1},
\end{equation*}
where $\alpha_i$, $\sigma_i$ are quantum numbers and $\circ$ represents the group product.
The most convenient representation for the MPS tensors is in a matrix-like form where the physical index $\sigma_i$ is fused together with one of the auxiliary indexes $\alpha_i$ or $\alpha_{i+1}$.
 We then profit from the fact that the matrix $A_{\sigma_i \alpha_i;\, \alpha_{i+1}}$ is block-diagonal in the quantum number labels, and we store only the dense blocks. Linear algebra operations acting on these dense matrices are dispatched to optimized BLAS and LAPACK libraries available on the user machine\footnote{The reader is redirected to the ALPS documentation~\cite{bauer2011-alps, alps-web} for the exact configuration options.}.

The MPS codes presented in this paper are implemented to make use of Abelian symmetries.
Dealing with conserved quantum numbers involves an intense calculation of indexes labels. We therefore set the symmetry at compile-time to allow for a maximum amount of compiler optimization. By default, the code is compiled for $U(1)^{\otimes n}$ symmetry, where the value $n$ is fixed during compilation by the macro \verb|DMRG_NUMSYMM|, as well as $\mathbb{Z}_2$ symmetry\footnote{$\zz$ symmetry is implemented only for some particular models. Refer to~\ref{app:z2 symm} for details on how to use it.}. Other symmetries may easily be added.

\subsection{Parallelization}
We parallelize our applications for shared memory architectures using the OpenMP compiler directives. The \verb|parallel for| loop directive is employed, for example, in the contraction of the local MPS tensor with the local MPO terms. In this routine several loops over the size of the auxiliary MPO dimension operate on independent parts. Since the MPO auxiliary dimension grows linearly with the width of a 2d system, we profit from more parallelism for large systems. The OpenMP parallelization is enabled at compile-time with the CMake option \verb|ALPS_ENABLE_OPENMP|, and can be controlled at run-time via standard environment variables such as \verb|OMP_NUM_THREADS|.

Many vendors provide support for multi-threaded BLAS and LAPACK libraries. For large matrices sizes our application will profit automatically from this parallelization, too. However, special care must be taken when using both the OpenMP parallelization and a parallel BLAS/LAPACK library simultaneously; this will only lead to improved scaling when using a compiler that supports such nested parallelism.

\section{Codes and examples}
\label{sec:code-example}

Our set of MPS tools is composed of an application for the calculation of ground states and low energy excitations (\verb|mps_optim|), a time evolution application (\verb|mps_evolve|) and two utilities (\verb|mps_meas| and \verb|mps_overlap|) to perform measurements on the wave functions generated with the previous programs.

The interface of the applications is consistent with other ALPS applications, i.e. the same simulation can easily be run with exact diagonalization, the MPS optimization or even Monte Carlo applications. On the outer level a scheduler reads an XML input file listing all parameter sets to be computed. These parameter files can be created with the standard ALPS tools: (1) converting text files to XML with the \verb|parameter2xml| command line tool, (2) using the convenient Python functions available in the \verb|pyalps| package or (3) generating a workflow from VisTrails~\cite{silva2007,vistrails}.

Each simulation generates three output files: \verb|simulation.out.h5| contains the simulation parameters, the iteration results and the final measurements in HDF5 format (see~\ref{app:hdf5 scheme} for the detailed HDF5 schemes); \verb|simulation.out.xml| contains the simulation parameters and the final results in XML format conforming with the ALPS schemas~\cite{bauer2011-alps, alps-web}; \verb|simulation.out.chkp| is the simulation checkpoint file containing the simulation status and the final MPS wave function that can be used to restore the same simulation or as input for other calculations.

Together with the applications, in the ancillary files, we ship full example simulations to illustrate how to create parameter files, run simulations and analyze the data. In sections \ref{sec:hubbard} and \ref{sec:domain wall} we describe some of these results.

\subsection{Input parameters}
In Table~\ref{tab:common params}, Table~\ref{tab:optim params} and Table~\ref{tab:evolve params} we list all parameters and the default values used by the applications. They are split between common MPS parameters (valid in all codes), optimization parameters (available only in \verb|mps_optim|) and time evolution specific parameters (for \verb|mps_evolve|). Models, lattice and measurements are parameters of the ALPS model and lattice libraries, we point the reader to the ALPS references~\citep{bauer2011-alps, alps-web} for further details.

\subsubsection{Initial states}
The MPS applications can be initialized with four different types of initial states.

\paragraph{default}
The default initial state constructs an MPS which contains all allowed quantum number sectors that fulfill the total quantum numbers chosen by the user. In each MPS tensor all sectors will have a maximum size of $5\times 5$ and they will be filled with random numbers.

\paragraph{thin}
The thin initial state is an extension of the default option. After constructing the default MPS the routine performs a compression retaining only a total $20$ truncated states per site. This initial state reduces considerably the number of sectors that the code has to tackle in the warm up sweeps. Sometimes it leads to faster convergence.

\paragraph{basis state}
A many body basis state is easily encoded in a product state as an MPS with bond dimension $M=1$. With this option the user can design its own initial state listing the desired local quantum numbers in the input file as illustrated in Table~\ref{tab:common params}.

\paragraph{input file}
The checkpoint file generated by a simulation can easily be reused in a new simulation by specifying it in the \texttt{initfile} parameter. This option is often used to start a time evolution from the ground state of an other model.

\subsubsection{Temporary memory}
An MPS simulation easily requires more than 100 GB of memory for the storage of boundary objects, i.e. partial contractions of the MPS network needed to reduce the complexity of the variational optimization to only $\mathcal{O}(M^3)$.

Our codes are capable of efficiently move this temporary memory to disk. This option is enabled by giving a non-empty value to the parameter {\tt storagedir}. Since the application will often write $>100$GB of temporary data it is advised to use fast filesystems, e.g. running on a local hard disk instead of a network mount. Once a simulation finishes the user can safely erase all temporary data.

\subsection{Example 1: Itinerant fermions in the Hubbard ladder}
\label{sec:hubbard}
As an example for the ground state search we show results for the pairing properties of the Hubbard model on a two-leg open ladder. The Hamiltonian of this model is
\begin{multline}
\hat H = -t \sum_{\langle ij \rangle, \sigma} \left[ \hat c^\dagger_\sigma(i) \hat c_\sigma(j) + \text{h.c.} \right] \\+ U \sum_i \hat n_\uparrow(i) \hat n_\downarrow(i)  , \quad \quad
\end{multline}
where $\hat c^\dagger_\uparrow(i)$ and $\hat c^\dagger_\downarrow(i)$ create a fermion at site $i$ with spin up and down, respectively, and $\hat n_\sigma(i) = \hat c^\dagger_\sigma(i) \hat c_\sigma(i)$ counts the number of particles with spin $\sigma \in \{\uparrow, \downarrow\}$ on site $i$. For a relatively strong repulsive interaction $U/t=8$ and an average filling $\overline n = 0.875$ this model is expected to fall into the Luther-Emery universality class~\cite{noack1996,luther1974} with a spin gap and a gapless charge excitation. To leading order both the long-range pairing correlation function $D(|i-j|) = D(l) \sim l^{-1/K_\rho}$ -- defined as
\begin{equation}
D(i,j) = \langle \Delta(i)\, \Delta^\dagger(j) \rangle
\end{equation}
with $\Delta^\dagger(i) = [ \hat c^\dagger_\uparrow(i,1) \hat c^\dagger_\downarrow(i,2) - \hat c^\dagger_\downarrow(i,1) \hat c^\dagger_\uparrow(i,2) ]$ creating a singlet on the $i$th-rung -- and the long-range charge density wave correlation function $C(|i-j|) = C(l) \sim l^{-K_\rho}$ -- defined as 
\begin{equation}
C(i,j) = \langle \hat n(i)\, \hat n(j) \rangle - \langle \hat n(i)\rangle \langle \hat n(j) \rangle
\end{equation}
 -- decay algebraically with an exponent that depends only on the non-universal parameter $K_\rho$, which we have to determine from the simulation at the chosen model parameters.

Long-range correlations have to propagate through the whole system and are therefore easily suppressed when truncating to only a few DMRG states. To treat the DMRG truncation correctly we run several instances of the simulation with increasing bond dimension $M$, until the observables converges or we can confidently extrapolate their value to $M \rightarrow \infty$.

\begin{figure}
  \centering
  \includegraphics[width=\columnwidth]{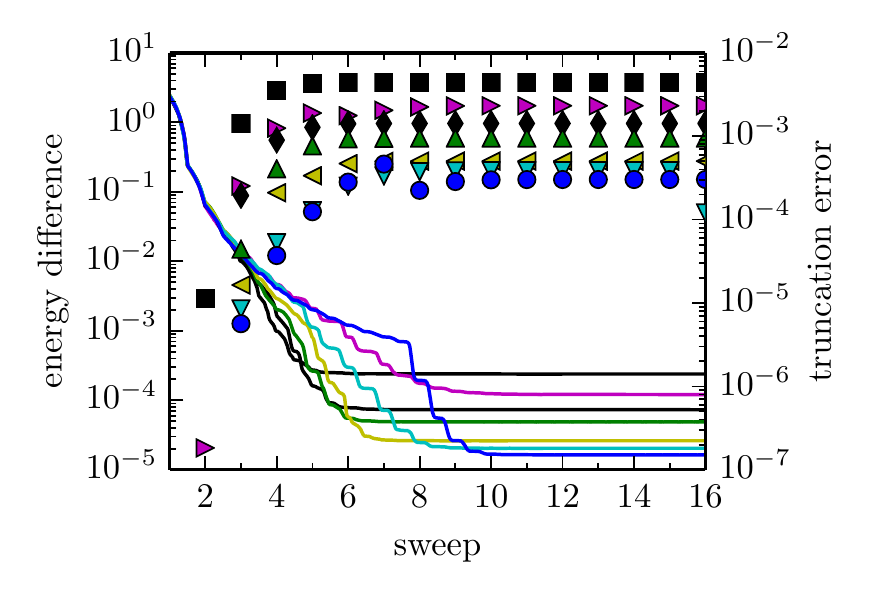}
  \caption{Iteration history of the energy convergence (solid lines) compared to the extrapolated final energy obtained in Figure~\ref{fig:hubbard energy extrapolation} and the corresponding truncation error (full points) for a $2\times 96$ system with average filling $\overline n = 0.875$ and $U/t = 8$. From top to bottom the data shown correspond to a maximum bond dimension $M = 800, 1200, 1600, 2000, 2800, 3200, 3600$.}
  \label{fig:hubbard iterations}
\end{figure}

In this example we simulate a $2\times 96$ ladder with bond dimension varying from a computationally cheap $M=800$ (convergence takes ca. 2.3h on 5 cores) to a computationally expensive $M=3600$ (convergence takes 46h on 5 cores). The run script is:
\begin{lstlisting}[style=pythonCode, numbers=left, basicstyle=\scriptsize]
parms = []
for M in [800, 1200, 1600, 2000, 2800, 3200, 3600]:
  p = dict()
  p['SWEEPS'    ] = 16
  p['MAXSTATES' ] = M
  p['init_state'] = 'thin'
  p['LATTICE'   ] = 'open ladder'
  p['L'         ] = 96
  p['MODEL_LIBRARY'] = 'mymodel.xml'
  p['MODEL'        ] = 'fermion Hubbard'
  p['t'            ] = 1
  p['U'            ] = 8
  p['CONSERVED_QUANTUMNUMBERS'] = 'Nup,Ndown'
  p['Nup_total'               ] = 84
  p['Ndown_total'             ] = 84
  p['MEASURE[EnergyVariance]'] = 1
  p['MEASURE_HALF_CORRELATIONS[pair field 1]'] = 'field_du:fielddag_ud'
  parms.append(p)

## write the input file and run the simulation
infiles=pyalps.writeInputFiles('sim',parms)
res=pyalps.runApplication('mps_optim',infiles)
\end{lstlisting}

\begin{figure}
  \centering
  \includegraphics[width=\columnwidth]{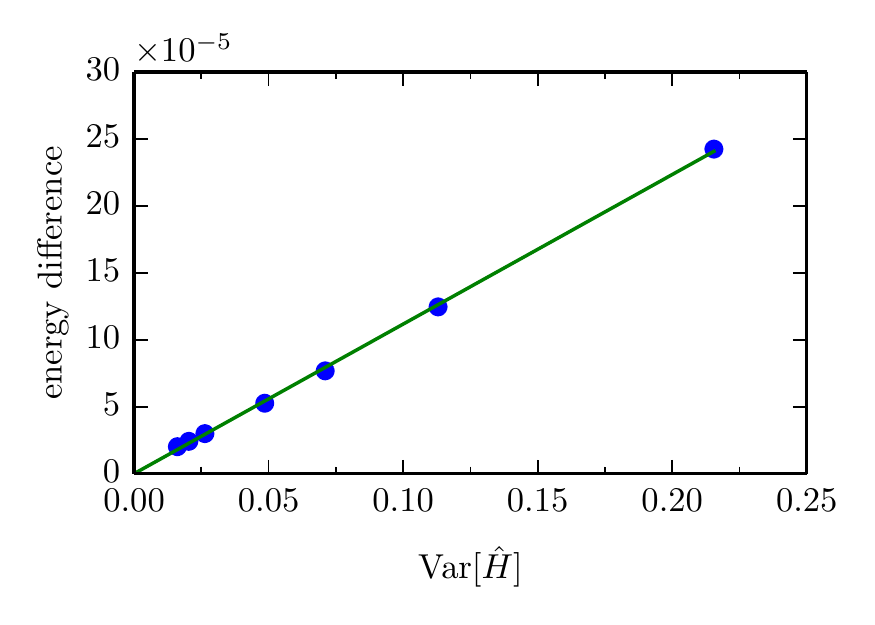}
  \caption{Extrapolation of the energy per particle as a function of the energy variance $\Var{\hat H} = \langle \hat H^2 \rangle - \langle \hat H \rangle^2$ for $2\times 96$ system with average filling $\overline n = 0.875$ and $U/t = 8$. The linear extrapolation returns the ground state energy $e_0=-0.72577\, t$ which has been subtracted from y-axis in the figure.}
  \label{fig:hubbard energy extrapolation}
\end{figure}

\begin{figure}
  \centering
  \includegraphics[width=\columnwidth]{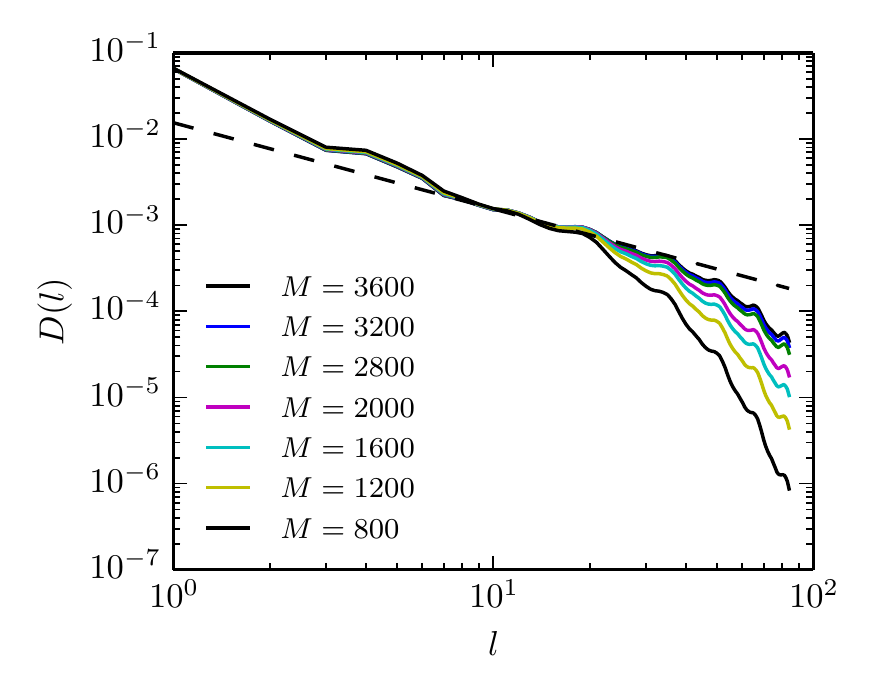}
  \caption{Rung-rung pairing correlation function $D(l)$ as a function of the distance between rungs for a $2\times 96$ system with average filling $\overline n = 0.875$ and $U/t = 8$. Values at distance $l$ are obtained by averaging 10 pairs $|i-j|=l$ around the center of the ladder. The dashed line is a reference algebraic decay with exponent $1$.}
  \label{fig:hubbard pairfield}
\end{figure}

Results of these simulations are reported in Figures~\ref{fig:hubbard iterations}, \ref{fig:hubbard energy extrapolation}, \ref{fig:hubbard pairfield}, where we illustrate the convergence history, the energy extrapolation and the decay of the pairing correlation function.

The iteration history of the simulation is loaded from the result file \verb|simulation.out.h5| with the Python code:
\begin{lstlisting}[style=pythonCode, numbers=left, basicstyle=\scriptsize]
resfiles = pyalps.getResultFiles(prefix='sim')
iters = pyalps.loadIterationMeasurements(resfiles, what=['Energy', 'TruncatedWeight'])
en_vs_iter = pyalps.collectXY(iters,
       x='iteration', y='Energy',
       foreach=['MAXSTATES'])
pyalps.plot.plot(en_vs_iter)
\end{lstlisting}
Because of the iterative optimization in DMRG, one should always check for convergence with the number of sweeps through the system. In Fig.~\ref{fig:hubbard iterations}, we see that both the ground state energy and the truncated weight have converged in our simulations.

Note that in the above parameters we enable the calculation of the energy variance $\Var{\hat H} = \langle \hat H^2 \rangle - \langle \hat H \rangle^2$. This is a very convenient quantity for extrapolating observables to $M\to \infty$, corresponding to $\Var{\hat H} \to 0$ for an eigenstate of $\hat H$. As an example the ground state energy depends to first order linearly on $\Var{\hat H}$. In Figure~\ref{fig:hubbard energy extrapolation} we show the extrapolation of the energy generated with the following code:

\begin{lstlisting}[style=pythonCode, numbers=left, basicstyle=\scriptsize]
## load data
resfiles = pyalps.getResultFiles(prefix='sim')
data = pyalps.loadEigenstateMeasurements(resfiles, ['Energy', 'EnergyVariance'])
en_vs_var = pyalps.ResultsToXY(data, 'EnergyVariance', 'Energy')

## linear fit
xgrid = np.linspace(0, max(en_vs_var[0].y))
coeff = np.polyfit(en_vs_var[0].x, en_vs_var[0].y, deg=1)

## plot data and fit
pyalps.plot.plot(en_vs_var)
plt.plot(xrgid, np.polyval(coeff,xgrid))
\end{lstlisting}

The evaluation of the pairing correlation function $D(i,j)$ includes terms like $\mathtt{field\_du : fielddag\_up} = \hat c_\downarrow(i,1) \hat c_\uparrow(i,2) \hat c^\dagger_\uparrow(j,1) \hat c^\dagger_\downarrow(j,2)$ which was listed as a correlation measurement in the parameters listed above. Since \verb|field_du| and \verb|fielddag_ud| are defined in the file \verb|mymodel.xml|~\footnote{The model description file mymodel.xml is available in the auxiliary files of this paper.} as bond operators, the application applies the four operators to all pairs of nearest neighbors sites, i.e. $i, i+1, j, j+i$ for all $i=1,\dots,N-4$ and $j=i+2,\dots,N$, effectively scaling as $\mathcal{O}(N^2)$.

Form the measured $D(i,j)$ we compute $D(l) = D(|i-j|)$ by averaging over a few rungs at distance $l=|i-j|$ around the center of the ladder. This technique averages out oscillations induced by the open boundaries. Results are shown in Figure~\ref{fig:hubbard pairfield}, where one notes a very slow convergence of the pairing correlation function $D(l)$ with the retained DMRG states; for $M=800$ pairing at large distances is underestimated. Large enough MPS matrices, like the $M=3600$ data, are needed to correctly analyze the system.

From this example simulation we observe that $D(l)$ decays with an exponent consistent with $K_\rho > 1$, which is the condition for a dominant superconducting phase in the Luther-Emery universality class.

\subsection{Example 2: Evolution of a domain wall}
\label{sec:domain wall}
As a second example we run a simple time evolution simulation of the Heisenberg spin XX model, described by the Hamiltonian
\begin{equation}
\label{eq:xx model}
\hat H = J \sum_i \left[ S^x(i) S^x(i+1) + S^y(i) S^y(i+1) \right],
\end{equation}
with $S^x(i)$, $S^y(i)$ the $x$, $y$ component of the spin$-1/2$ operator $\vec S(i)$.

The time evolution of an initial domain wall $\ket{\psi(t=0)} = \ket{\downarrow\downarrow \dots \downarrow \uparrow \dots \uparrow\uparrow}$ -- a state with all up spins on the right side of the chain and the down spins on the left -- under the spin XX Hamiltonian was solved exactly by a Jordan-Wigner transformation to free fermions in Ref.~\cite{antal1999}. In particular, the local magnetization at distance $n$ from the center after time $t$ is described by a sum of Bessel functions as
\begin{equation}
\label{eq:domain wall mag}
M(n,t)=-\frac{1}{2}\sum_{i=1-n}^{n-1}j_i^2(t),
\end{equation}
with $j_i(t)$ is the Bessel function of order $i$.

\begin{figure}
  \centering
  \includegraphics[width=\columnwidth]{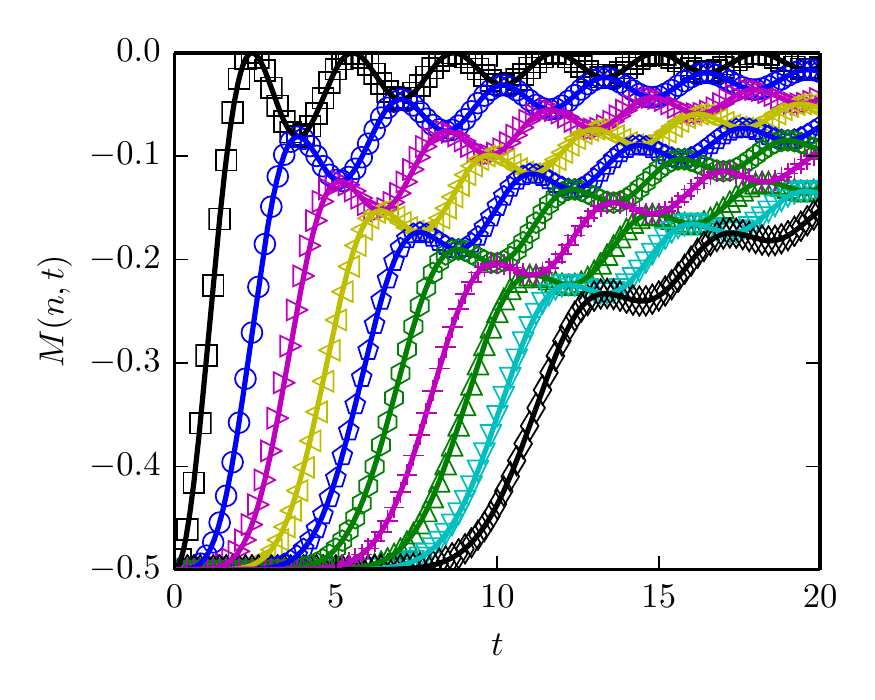}
  \caption{Magnetization $M(n,t) = \langle \psi(t) | \hat S_z(L/2 - n) | \psi(t) \rangle$ at time $t$ and distance $n$ from the middle of the chain. Points are simulation results with $n=1 - 10$ (from top to bottom), solid lines are the corresponding exact solutions computed from Eqn.~\ref{eq:domain wall mag}.}
  \label{fig:domain wall magnetization}
\end{figure}

The setup script is very similar to the one previously used for the ground state calculations, for a simulation of 500 time steps up to a total time $\tau = 20 \hbar / J$ it reads:
\begin{lstlisting}[style=pythonCode, numbers=left, basicstyle=\scriptsize]
total_time = 20
nsweeps    = 500
p['MAXSTATES'       ] = 40
p['TIMESTEPS'       ] = nsteps
p['DT'              ] = nsteps/total_time
p['measure_each'    ] = 5
p['init_state'      ] = 'local_quantumnumbers'
p['initial_local_Sz'] = ','.join(  ['-0.5']*25
                                 + ['0.5']*25)
p['ALWAYS_MEASURE'  ] = 'Local Magnetization'
p['LATTICE'         ] = 'open chain lattice'
p['L'               ] = 50
p['MODEL'           ] = 'spin'
p['Jxy'             ] = 1
p['CONSERVED_QUANTUMNUMBERS'] = 'Sz'
p['Sz_total'                ] = 0
p['MEASURE_LOCAL[Local Magnetization]'] = 'Sz'

## write the input file and run the simulation
infiles=pyalps.writeInputFiles('sim',parms)
res=pyalps.runApplication('mps_evolve',infiles)
\end{lstlisting}
Note the definition of the initial state {\tt initial\_local\_Sz} listing the value of the $S^z$ quantum number on each sites.

In Figure~\ref{fig:domain wall magnetization} the local magnetization calculated with the above simulation is compared with the exact solution from Eqn.~\ref{eq:domain wall mag} showing very good agreement. To produce a similar figure one simply has to load the results for each time step with the \verb|loadIterationMeasurements| function and select the magnetization at distance $n$ from the center, as shown in the following snippet.
\begin{lstlisting}[style=pythonCode, numbers=left, basicstyle=\scriptsize]
## simulation results
resfiles = pyalps.getResultFiles(prefix='sim')
data = pyalps.loadIterationMeasurements(resfiles, what=['Local Magnetization'])
## select the Local Magnetization at distance `loc` from the center
numeric_mag = []
for d in pyalps.flatten(data):
  L = d.props['L']
  for loc in range(1, 11):
    q = pyalps.DataSet()
    q.props = deepcopy(d.props)
    q.props['loc'] = loc
    q.y = [ q.y[0][L/2-loc] ]
    numeric_mag.append(q)
## plot Magnetization as a function of time
mag_vs_time = pyalps.collectXY(numeric_mag,
      x='Time', y='Local Magnetization',
      foreach=['loc'])
pyalps.plot.plot(mag_vs_time)
\end{lstlisting}

A slightly more complex version generating a movie of the local magnetization per site is also provided in the source code of the example.

\section{Acknowledgments}
We gratefully acknowledge support by the wider ALPS community. We are indebted to
Philippe Corboz and Ulrich Schollw\"ock for useful discussions in the design process,
and to U. Schollw\"ock for providing a draft of his Python code. We also thank
Jan Gukelberger, Hiroshi Shinaoka and Lei Wang for being early users of the framework and applications.
Miles Stoudenmire provided an ad-hoc ITensor~\cite{itensor} application for our benchmarks.
Simulations were performed on the PASC M\"onch cluster at ETH Zurich.

\begin{table*}[h]
\begin{center}
\caption{Common simulation parameters}
\begin{tabular}{l l p{.54\textwidth}}
\toprule
{\bf Parameter}   & {\bf Default}  & {\bf Description}  \\ \midrule
{\tt MAXSTATES}   &  & Maximum size of the matrices $M_{\sigma_i}$ \\
{\tt TRUNCATION}  & 1e-8 & Smallest singular value to be kept \\
{\tt seed}  & 42 & Seed of the internal random number generator \\
{\tt CONSERVED\_QUANTUMNUMBERS}  & & Comma-separated list of the quantum numbers to be conserved \\
{\tt \textit{QN}\_total}  & & Total value of the quantum number \textit{QN}. \textit{QN} is one of the quantum numbers defined in the parameter {\tt CONSERVED\_QUANTUMNUMBERS}. Example: {\tt N\_total = 8} fixes the total number of particles to 8.  \\
{\tt initfile}  &  & Path to an existing checkpoint MPS to use it as initial MPS \\
{\tt init\_state}  & default & Initial state of the simulation (in case \texttt{initfile} is not set). Possible values: `default', `thin', `local\_quantumnumbers'. \\
{\tt initial\_local\_\textit{QN}}  & & Comma-separated list with the value of the local quantum number \textit{QN} at every site \\
{\tt MEASURE\_LOCAL[\textit{NAME}]}  &  & Defines a new local measurement called \textit{NAME}. Its value specifies the operator to be measured.  Example: {\tt MEASURE\_LOCAL[Local density] = "n"} \\
{\tt MEASURE\_AVERAGE[\textit{NAME}]}  &  & Defines a new average measurement called \textit{NAME}. Its value specifies the operator to be measured.  Example: {\tt MEASURE\_AVERAGE[Density] = "n"} \\
{\tt MEASURE\_CORRELATIONS[\textit{NAME}]}  &  & Defines a new correlation measurement called \textit{NAME}. Its value specifies the operators to be correlated separated by ":" (colon) If the operators define two or more \verb|BONDOPERATORS|, e.g., \verb|Op_A| and \verb|Op_B| the correlation measurement compute all observables $\matrixel{\psi}{\mathtt{Op_A}(i, i+1)\, \mathtt{Op_B}(j, j+1)}{\psi}$. Example: {\tt MEASURE\_CORRELATIONS[Onebody Density Matrix] = "bdag:b"} \\
{\tt MEASURE\_HALF\_CORRELATIONS[\textit{NAME}]}  &  & Same as {\tt MEASURE\_CORRELATIONS} but it does not exchange the order of operators. If the input is, e.g., {\tt bdag:b}, the first operator will be evaluated at all locatios $i\in[0,L-2]$ but the second operator only at locations $j\in[i,L-1]$. \\
{\tt MEASURE\_LOCAL\_AT[\textit{NAME}]}  &  & Syntax for the value: {\tt "op\_1:\dots:op\_n | $(i_{1\,1},\dots,i_{1\,n})$, $(i_{2\,1},\dots,i_{2\,n})$, \dots"}.  Defines a new measurement called \textit{NAME} where the sequence of operators {\tt op\_1:\dots:op\_n} is applied to all tuples of indices (of length $n$ like the number of operators) listed after the vertical bar symbol "|".  \\
{\tt MEASURE[EnergyVariance]}  & False & Measure the energy variance \\
{\tt MEASURE[Entropy]}  & False & Measure the von Neumann entropy \\
{\tt MEASURE[Renyi2]}  & False & Measure $n=2$ Renyi entropy \\
{\tt ALWAYS\_MEASURE}  &  & Comma-separated list of measurements to evaluate at the end of every sweep \\
{\tt COMPLEX}  & False & Use complex numbers. For time evolution simulations the default value is changed to True. \\
{\tt storagedir}  & & Path to the directory to be used for temporary storage. If empty, temporary storage is disabled. \\
\bottomrule
\end{tabular}
\label{tab:common params}
\end{center}
\end{table*}

\begin{table*}[h]
\begin{center}
\caption{Additional parameters for `mps\_optim'}
\begin{tabular}{l l p{.5\textwidth}}
\toprule
{\bf Parameter}   & {\bf Default}  & {\bf Description}  \\ \midrule
{\tt SWEEPS}  & & Number of sweeps \\
{\tt NUMBER\_EIGENVALUES}  & 1 & Number of eigenstate to target \\
{\tt optimization}  & twosite & Optimisation algorithm. Possible values are `singlesite', `twosite'. \\
{\tt ngrowsweeps}  &  &  Single site only. Number of initial sweeps where the correction factor has the value {\tt alpha\_initial}. \\
{\tt nmainsweeps}  &  &  Single site only. Number of sweeps after {\tt ngrowsweeps} where the correction factor has the value {\tt alpha\_main}. \\
{\tt alpha\_initial}  & 1e-2 & Correction factor for single site optimization~\cite{white2005} in the initial part \\
{\tt alpha\_main}  & 1e-4 & Correction factor for single site optimization~\cite{white2005} in the main part \\
{\tt alpha\_final}  & 1e-8 & Correction factor for single site optimization~\cite{white2005} in the final part \\
{\tt ietl\_jcd\_toll}  & 1e-8 & Convergence tolerance of the Jacobi-Davidson solver \\
{\tt ietl\_jcd\_gmres}  & 0 & Convergence tolerance of the Jacobi-Davidson solver \\
{\tt ietl\_jcd\_maxiter}  & 8 & Maximum number of iterations in the Jacobi-Davidson solver \\
\bottomrule
\end{tabular}
\label{tab:optim params}
\end{center}
\end{table*}

\begin{table*}[h]
\begin{center}
\caption{Additional parameters for 'mps\_evolve'}
\begin{tabular}{l l p{.5\textwidth}}
\toprule
{\bf Parameter}   & {\bf Default}  & {\bf Description}  \\ \midrule
{\tt DT}  &  & Time step \\
{\tt IMG\_TIMESTEPS}   & 0 & Number of imaginary time sweeps to perform before the proceeding with the real time evolution \\
{\tt TIMESTEPS}   &  & Total number of sweeps \\
{\tt te\_order}  & fourth & Order of the trotter decomposition. Possible values: `second', `fourth'. \\
{\tt te\_type}  & nearest neighbors & Type of time evolution algorithm. Possible values: `nearest neighbors', `mpo'. \\
{\tt chkp\_each}  & 1 & A checkpoint is created every {\tt chkp\_each} time steps. \\
{\tt measure\_each}  & 1 & Measurements are performed every {\tt measure\_each} time steps. \\
{\tt update\_each}  & -1 & Update the Hamiltonian parameters every {\tt update\_each} time steps. Used while quenching Hamiltonian parameters. A negative value will never change the Hamiltonian. \\
{\tt \textit{P}[Time]}  & & Values assigned to the parameter \textit{P} at each time step \\
\bottomrule
\end{tabular}
\label{tab:evolve params}
\end{center}
\end{table*}

\appendix

\section{Algorithms for time evolution}
\subsection{Higher order Suzuki-Trotter decomposition}
\label{app:trotter}
The time evolution unitary operator for a small time step $\exp(-i \hat H \Delta t)$ is conveniently transformed according to the Suzuki-Trotter decomposition. In the \verb|mps_evolve| application the user can choose between a second or fourth order expansion with the parameter \verb|te_order=second| or \verb|te_order=fourth|, respectively.

The second order expands the unitary time evolution operator up to an error $\mathcal{O}(\Delta t^3)$, namely
\begin{equation}
  \label{eq:trotter 2nd}
  \exp(-i \hat H \Delta t) = S\left(\frac{\Delta t}{2}\right) + \mathcal{O}(\Delta t^3),
\end{equation}
with $S(t)$ defined as
\begin{equation*}
  S(\Delta t) =  \left( \prod_{n=1}^{N} e^{-i \hat h_n \Delta t} \right) \cdot \left( \prod_{n=N}^{1} e^{-i \hat h_n \Delta t} \right),
\end{equation*}
where the Hamiltonian operator $\hat H$ has been split into $N$ non-commuting terms $\hat h_n$.

Fourth order expansion provides a more accurate time step $\mathcal{O}(\Delta t^5)$ at the expense of applying more operators to the MPS wave function. The decomposition reads
\begin{equation}
  \label{eq:trotter 4th}
  \exp(-i \hat H \Delta t) = \prod_{j=1}^5 S\left(p_j \frac{\Delta t}{2}\right) + \mathcal{O}(\Delta t^5)
\end{equation}
with $p_1 = p_2 = p_4 = p_5 = p = \frac{1}{4 - 4^{1/3}}$ and $p_3 = 1 - 4 p$.

\subsection{Exponential of fermionic bond terms}
\label{app:fermion exponential}
According to the Jordan-Wigner transformation an operator $\hat O(i,j) = \hat c_i \otimes \hat c_j$ acting on the bond between site $i$ and $j$ with the local operators $\hat c_i$ and $\hat c_j$, respectively, is decomposed into a chain of bosonic operators $\hat b_i$ and $\hat b_j$ with filling sign matrices $\hat f_k$ acting on sites $i<k<j$.

The exponential of this operator can be split into tensor product of exponential as
\begin{align}
  e^{\alpha \hat O(i,j)} =& \sum_n \frac{\alpha^n}{n!} \hat b_i^n \otimes \hat f_{i+1}^n \otimes \cdots \otimes \hat b_i^n \notag \\
  =& \quad \sum_{n \in \text{even}} \frac{\alpha^n}{n!} \hat b_i^n \otimes  \mathbb{\hat I}_{i+1} \otimes \cdots \otimes \hat b_i^n \notag \\
  & + \sum_{n \in \text{odd}} \frac{\alpha^n}{n!} \hat b_i^n \otimes \hat f_{i+1} \otimes \cdots \otimes \hat b_i^n  \notag \\
  =& \frac{1}{2} \left( B(\alpha) + F(\alpha) \right) + \frac{1}{2} \left( B(-\alpha) - F(-\alpha) \right),
\end{align}
with
\begin{align*}
  B(\alpha) =& e^{\alpha \hat b_i} \otimes e^{\alpha \mathbb{\hat I}_{i+1}} \otimes \cdots \otimes e^{\alpha \hat b_j} \\
  F(\alpha) =& e^{\alpha \hat b_i} \otimes e^{\alpha \hat f_{i+1}} \otimes \cdots \otimes e^{\alpha \hat b_j}.
\end{align*}

\section{Extensions for $\zz$ symmetry}
\label{app:z2 symm}

The additional $\zz$ symmetry, which is useful in the simulation of systems such as superconductors,
can be enabled at compile-time using the CMake option \verb|DMRG_BUILD_SYMMETRIES|,
and setting \verb|symmetry = "Z2"| in the parameter file. In Listing~\ref{tsc}, we provide the skeleton
of a model definition for a superconductor, where the $\zz$ symmetry is fermion parity, i.e.
particle number modulo 2. The full example, along with a script that shows how to detect the
topological phase transition in the Kitaev wire model~\cite{kitaev2001}, is included in the auxiliary
files to this manuscript.

\begin{lstlisting}[style=listXML, float=*,basicstyle=\footnotesize, caption={Superconductor with $\zz$ symmetry. \label{tsc}}]
<MODELS>
    <SITEBASIS name="spinless fermion">
        <QUANTUMNUMBER name="P" min="0" max="1" type="fermionic"/>
        <OPERATOR name="c" matrixelement="1">
            <CHANGE quantumnumber="P" change="1"/>
        </OPERATOR>
        <!-- analogously define cdag -->
    </SITEBASIS>
    
    <!-- define BASIS -->
    <!-- define bond operators -->
    
    <HAMILTONIAN name="tsc">
        <BASIS ref="spinless fermion"/>
        <BONDTERM source="i" target="j">
            -t*fermion_hop(i,j) + D*pairing(i,j)
        </BONDTERM>
    </HAMILTONIAN>
     
</MODELS>
\end{lstlisting}

\section{HDF5 schema}
\label{app:hdf5 scheme}

Simulation results are stored in the widely used Hierarchical Data Format (HDF). In this section we list the schemes used to store the simulation parameters, the measurement results and the iteration information. The ALPS Python module \verb|pyalps| provides helper functions which load the data automatically.

\subsection*{Parameters}
\begin{description}
\item[\path{/parameters/PARAMETERNAME}] \hfill\\
Value of the parameter \texttt{\textit{PARAMETERNAME}}.
\end{description}

\noindent Helper function:\\
\verb|pyalps.loadProperties(files)|

\subsection*{Measurements}
\begin{description}
\item[\path{/spectrum/results/OBSNAME/mean/value}] \hfill \\
Vector of expectation values of the observable \texttt{\textit{OBSNAME}}, each entry in the vector is a different eigenstate.

\item[\path{/spectrum/results/OBSNAME/labels}] \hfill \\
Labels for the x-coordinate of the observable.
\end{description}

\noindent Helper function: \\
\verb|pyalps.loadEigenstateMeasurements(files)|

\subsection*{Iterations}
\begin{description}
\item \hspace*{-\labelsep}%
    \path{/spectrum/iteration/NUM/parameters/PARAMETERNAME} \par
Value of the parameter \texttt{\textit{PARAMETERNAME}} specific to the iteration \texttt{\textit{NUM}}. Only modified parameters appear in this group.

\item \hspace*{-\labelsep}%
  \path{/spectrum/iteration/NUM/results/OBSNAME/mean/value} \par
Expectation values of the observable \texttt{\textit{OBSNAME}} at the iteration \texttt{\textit{NUM}}.

\item \hspace*{-\labelsep}%
 \path{/spectrum/iteration/NUM/results/OBSNAME/labels} \par
Labels for the x-coordinate of the observable.
\end{description}

\noindent Helper function: \\
\verb|pyalps.loadIterations(files)|

\section*{References}
\bibliographystyle{elsarticle-num}
\bibliography{refs.bib}

\end{document}